\documentclass[conference]{IEEEtran}
\IEEEoverridecommandlockouts
\usepackage{fancyhdr}
\usepackage[T1]{fontenc}
\usepackage{cite}
\usepackage{amsmath,amssymb,amsfonts}
\usepackage{algorithmic}
\usepackage{graphicx}
\usepackage{textcomp}
\usepackage{url}
\usepackage{xcolor}
\usepackage{listings}

\definecolor{CodeDarkRed}{HTML}{9D0006}
\definecolor{CodeGreen}{HTML}{79740E}
\definecolor{CodeOrange}{HTML}{AF3A03}
\definecolor{CodeBlue}{HTML}{076678}
\definecolor{CodePurple}{HTML}{8F3F71}
\definecolor{CodeAnnotation}{HTML}{2455A4}
\definecolor{CodeAqua}{HTML}{427B58}
\definecolor{CodeGray}{HTML}{928374}
\definecolor{CodeFg}{HTML}{3C3836}
\definecolor{CodeBg}{HTML}{FDFBEF}

\lstdefinelanguage{MLIR}{
  sensitive=true,
  alsoletter={\%.\#_},
  morecomment=[l]{//},
  morecomment=[s]{/*}{*/},
  morestring=[b]",
  morekeywords=[2]{
    tileplan.logical_read,tileplan.persistent_l1,tileplan.converted_view
  },
  morekeywords=[3]{
    tileflow.transfer,tileflow.local_read,tileflow.frame_delivery,
    tileflow.roll,tileflow.transpose,tileflow.broadcast,
    tileflow.masked_overlay,tileflow.row_copy,
    tileflow.col_mask,tileflow.row_mask,tileflow.mask_fill
  },
  morekeywords=[4]{
    task.group,task.epoch_acquire,task.epoch_ready,task.epoch_wait,task.epoch_free
  },
  morekeywords=[5]{d2mtile.unary,d2mtile.binary},
  morekeywords=[6]{shard_for,temporal_for,make_pivot_col,update},
  morekeywords=[7]{
    bf16,f16,f32,f64,index,memref,tensor,true,false,unit,
    axis,by,kind,from,active,policy,name,row,col,scalar
  },
  morekeywords=[8]{
    tileflow.realization_engine,tileflow.predication_target,
    tileflow.predication,tileflow.realization_basis,
    tileflow.conversion_site,task.streaming_granularity
  }
}

\lstdefinestyle{code-druvbox-light-high-small}{
  basicstyle=\ttfamily\scriptsize\color{CodeFg},
  backgroundcolor=\color{CodeBg},
  commentstyle=\itshape\color{CodeGray},
  stringstyle=\color{CodeAqua},
  columns=fullflexible,
  keepspaces=true,
  showstringspaces=false,
  upquote=true,
  tabsize=2,
  numbers=left,
  numberstyle=\tiny\color{CodeGray},
  numbersep=5pt,
  frame=lines,
  breaklines=true,
  breakatwhitespace=false,
  prebreak=\raisebox{0ex}[0ex][0ex]{\ensuremath{\hookleftarrow}},
  aboveskip=4pt,
  belowskip=4pt,
  captionpos=b
}

\lstdefinestyle{mlir-druvbox-light-high-small}{
  style=code-druvbox-light-high-small,
  language=MLIR,
  keywordstyle=[2]\bfseries\color{CodeDarkRed},
  keywordstyle=[3]\bfseries\color{CodeOrange},
  keywordstyle=[4]\bfseries\color{CodeGreen},
  keywordstyle=[5]\bfseries\color{CodeBlue},
  keywordstyle=[6]\bfseries\color{CodePurple},
  keywordstyle=[7]\bfseries\color{CodeAqua},
  keywordstyle=[8]\bfseries\color{CodeAnnotation},
  literate=
    {<row>}{{\bfseries\color{CodeAqua}<row>}}5
    {<col>}{{\bfseries\color{CodeAqua}<col>}}5
    {<scalar>}{{\bfseries\color{CodeAqua}<scalar>}}8
}

\def\BibTeX{{\rm B\kern-.05em{\sc i\kern-.025em b}\kern-.08em
    T\kern-.1667em\lower.7ex\hbox{E}\kern-.125emX}}
\usepackage[hidelinks,breaklinks=true]{hyperref}

\begin{document}

\title{Annotation-Driven Migration of CUDA Programs\\ to Tenstorrent Blackhole
}

\author{\IEEEauthorblockN{Ayumi Ohno}\IEEEauthorblockA{The University of Tokyo, Tokyo, Japan\\ayumi0130ohno@is.s.u-tokyo.ac.jp}\and
\IEEEauthorblockN{Shinya Takamaeda-Yamazaki}
\IEEEauthorblockA{
The University of Tokyo and RIKEN, Tokyo, Japan\\
shinya@is.s.u-tokyo.ac.jp
}
}

\maketitle
\thispagestyle{fancy}
\lhead{}
\rhead{}
\chead{}
\rfoot{}
\cfoot{}
\renewcommand{\headrulewidth}{0pt}
\renewcommand{\footrulewidth}{0pt}

\begin{abstract}
Tenstorrent Blackhole combines distributed local memories, explicit inter-core communication, and Tensix cores decoupling data movement from computation. CUDA offers a substantial HPC software base but leaves physical data placement and scheduling largely implicit. Migrating CUDA HPC kernels to Blackhole requires spatial mapping across cores and per-core coordination of compute and data-movement kernels.
We present an MLIR-based compiler deriving data placement, inter-core communication, and tile computation from a statically shaped affine CUDA subset. Declarative annotations express choices not fixed by the source, including compute/DM operation placement and streaming granularity.
We evaluate it on Gaussian elimination, five-point stencil, and symmetric rank-$k$ update. Relative to our compiler's default realizations, the best measured configurations achieve a $4.2\times$ speedup for BF16 Gaussian and $2.1$--$2.2\times$ for FP32 Gaussian and Jacobi. A choice's performance impact can reverse with surrounding policies, precision, and loop schedule, motivating comparison of alternative per-core realizations rather than independent policy selection.
\end{abstract}


\section{Introduction}
\label{sec:intro}

Spatial accelerators have emerged as hardware platforms for data-intensive computation, combining distributed local memories with explicit on-chip communication~\cite{chen2016eyeriss}.
Cerebras wafer-scale engines, AMD Versal AI Engines, and Tenstorrent Tensix core arrays are prominent examples~\cite{lie2023cerebras,gaide2019versal,vasiljevic2024blackhole}.
Beyond AI, such systems have also been explored for HPC workloads, including stencils and FFTs~\cite{rocki2020fast,orenes2023wafer,brown2023exploring,brown2024accelerating,piarulli2026stencil,taylor2026numerical,brown2025exploring}.
Tenstorrent Blackhole presents an additional mapping challenge: each Tensix core decouples data movement from computation~\cite{vasiljevic2024blackhole}.
Separate RISC-V processors execute compute and data-movement (DM) kernels and communicate through circular buffers (CBs) in L1 memory.
Mapping data and computation across Tensix cores is therefore only part of the problem: the compiler must also divide operations between the compute and DM kernels and coordinate their execution within each core.
We refer to this intra-core mapping problem as \emph{per-core realization}, a central focus of this work.

Tenstorrent's standard TTIR-to-TTNN path represents computation through tensor-level operations but does not expose the low-level data movement and synchronization choices used in optimized HPC kernels.
Prior Tenstorrent HPC studies~\cite{brown2024accelerating,piarulli2026stencil,taylor2026numerical,brown2025exploring} optimize kernels such as stencils and FFTs in TT-Metalium, rather than provide a general compilation path for such programs.
At a higher level, TileLoom~\cite{li2026tileloom} automates spatiotemporal mapping and inter-core dataflow for workloads such as FlashAttention and GEMM written in tile-based languages.
These efforts leave open a general compilation path that exposes alternative per-core realizations for HPC workloads with varied affine access and computation patterns.

A substantial body of HPC kernels and benchmark suites is available in CUDA~\cite{che2009rodinia,grauer2012auto}.
CUDA's SIMT abstraction describes computation and memory accesses over threads while leaving physical data placement, movement, and scheduling implicit.
Migrating these programs to Blackhole will improve the programming experience, but requires deriving both spatial tile dataflow and a decoupled per-core realization.

We present an MLIR-based compiler that performs this migration for an affine subset of CUDA whose per-core working sets fit in local L1.
Given affine iteration domains and access maps together with a compact placement specification, the compiler assigns data and computation to the core grid and derives the communication required for each access.
Within each core, it lowers thread-level computation to tile operations, constructs their operands using within-tile transformations, and converts guards to tile masks.
It then partitions the operations between the compute and DM kernels and coordinates their execution using L1 buffers and CB synchronization.

The key technical challenge in this migration is that source semantics do not determine a unique per-core realization.
For example, arithmetic may run in a compute or DM kernel; a guard may become a compute-side mask or a DM-side copy; work may be divided between the two DM kernels; and CB synchronization may operate at different granularities.
The performance of each choice depends on the resulting balance of compute and DM work, so assigning a fixed cost to each choice can be insufficient.
We expose these choices as declarative annotations.
The compiler checks their applicability, applies the corresponding IR transformations, and lowers the result without handwritten low-level kernels.

We evaluate Gaussian elimination, five-point stencil, and symmetric rank-$k$ update using CUDA programs derived from benchmark suites, subject to the scope and target-independent canonicalization of Section~\ref{sec:bg-workload}; for each workload, all configurations share the same CUDA input and differ only in their annotation sets.
The best measured annotation sets yield $4.2\times$ at BF16 for \texttt{gaussian} and $2.1$--$2.2\times$ at FP32 for \texttt{gaussian} and \texttt{jacobi\_2d}, relative to the compiler's own default realizations, in which arithmetic runs on compute, data manipulation on DM, guards become compute-side masked selects, and all DM work uses a single lane.
The preferred per-core realization depends on the surrounding policies, arithmetic precision, and loop schedule.

\section{Background}
\label{sec:bg}

\subsection{Tenstorrent Blackhole Architecture}
\label{sec:bg-hw}

\begin{figure}[htb]
  \centering
  \includegraphics[width=.79\columnwidth]{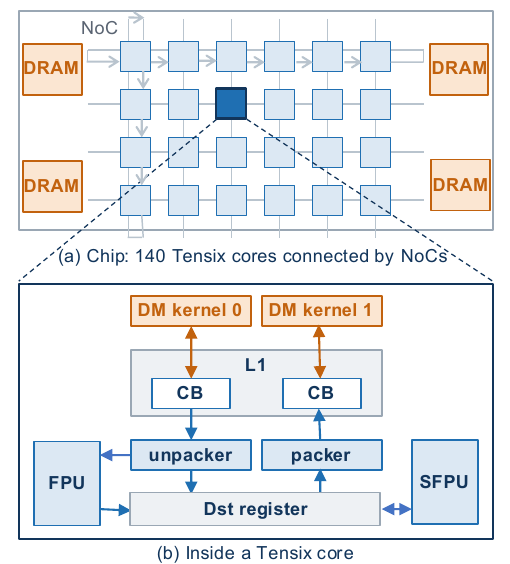}
  \caption{Blackhole chip and Tensix core organization.
  The two DM processors access the NoC interfaces and exchange data with compute kernels through CBs in L1.}
  \label{fig:tensix-overview}
\end{figure}

Blackhole is a spatial architecture that integrates 140 Tensix cores on a chip, of which our evaluated device exposes a $13\times10$ logical worker grid to TT-Metalium. Each core has 1.5~MiB of L1 SRAM.
Figure~\ref{fig:tensix-overview} shows the execution organization of a Tensix core.
Two RISC-V DM processors issue transfers to DRAM and remote cores through the two NoC interfaces~\cite{vasiljevic2024blackhole}.
In TT-Metalium, DM and compute kernels communicate through CBs in L1 memory.
The CBs stage operands and results and provide producer--consumer synchronization, allowing data transfer and computation to overlap.
We refer to DM0, DM1, and the compute pipeline as \emph{engines}.

Computation on a Tensix core is organized over $32 \times 32$-element tiles, each arranged as four $16 \times 16$-element faces.
The unpacker loads tiles from input CBs into source registers, the FPU computes from these registers into Dst registers, and the packer writes Dst tiles to an output CB.
The SFPU instead performs operations on values already staged in Dst.
For BF16 binary operations, the FPU is used when at least one operand is unpacked from CBs and the SFPU when both already reside in Dst. FP32 element-wise operations use the SFPU.

\subsection{Compiler Infrastructure}
\label{sec:bg-pipe}

We use Polygeist~\cite{moses2021polygeist} to translate CUDA programs into MLIR~\cite{lattner2021mlir}, enabling its affine-raising pass so that eligible loops, conditions, and array accesses are represented in MLIR's Affine Dialect.
Our compiler ultimately emits the \texttt{D2M} (Direct-to-Metal) Dialect of \texttt{tt-mlir} rather than the standard TTIR-to-TTNN path, because \texttt{D2M} can express computation over sharded tensors and core grids together with explicit engine assignment, data movement, and synchronization.

\subsection{Target Workloads and Scope}
\label{sec:bg-workload}

We evaluate the compiler on three CUDA workloads: \texttt{gaussian} from Rodinia~\cite{che2009rodinia} and \texttt{jacobi\_2d} and \texttt{syrk} from PolyBench/ACC~\cite{grauer2012auto}.
\texttt{gaussian} performs the guarded elimination update $A_{ij} \leftarrow A_{ij} - A_{it}/A_{tt} \cdot A_{tj}$ over $i>t$ and $j\ge t$.
Mapping this update requires inter-core transfers and within-tile broadcasts to align the pivot-dependent operands while preserving the update domain.
\texttt{jacobi\_2d} repeatedly applies a five-point stencil to interior points, leaving the global array boundary unchanged.
The stencil operands are constructed using within-tile shifts and tile-boundary fills from adjacent tiles, with inter-core transfers when those tiles reside on neighboring cores.
\texttt{syrk} computes $C \leftarrow \alpha A A^T + \beta C$.
Its reduction is realized as tile matrix multiplications with accumulated partial results and can be blocked to expose additional reuse.

We use the \texttt{jacobi\_2d} kernel body without restructuring, but manually adapt the other two inputs to limitations of the current frontend. 
For \texttt{gaussian}, we omit the update of the right-hand-side vector $b$ because our current frontend does not support distributing rank-1 arrays.
For \texttt{syrk}, we rewrite PolyBench's in-place reduction using a scalar accumulator followed by a single final update, $C_{ij}=\alpha\,acc+\beta C_{ij}$.
The current frontend requires this target-independent canonicalization, which can change floating-point rounding.
For all three workloads, we set the CUDA thread-block size to $32\times32$, because the current lowering maps the threads of one block to the elements of one Tensix tile. 
These adaptations add no workload-specific logic to the lowering and can be eliminated through future extensions (Section~\ref{sec:pipe-limitations}).
All three are evaluated at problem sizes whose per-core working sets remain within each core's 1.5~MiB L1, so that the measurements are not dominated by DRAM bandwidth.

\subsection{Prior Work on HPC Acceleration for Tenstorrent}
\label{sec:bg-related-tt-hpc}

Prior Tenstorrent stencil studies expose different data-movement bottlenecks depending on where data is kept between iterations.
Brown and Barton~\cite{brown2024accelerating} stream a 2-D \texttt{jacobi\_2d} stencil through device DRAM on Grayskull.
Their optimizations favor fewer, larger, contiguous DRAM requests and reduce redundant local data copying.
Piarulli and De Sensi~\cite{piarulli2026stencil} instead use heterogeneous CPU--Wormhole implementations of a 2-D five-point stencil.
The CPU constructs shifted submatrices or a stencil-to-row layout, and data crosses PCIe on every iteration, making CPU-side preprocessing and transfer overhead central to their analysis.

Taylor et al.~\cite{taylor2026numerical} keep intermediate values in per-core L1 rather than DRAM and implement a 3-D seven-point stencil with halo exchange on Wormhole.
Neighboring elements are aligned through tile shifts; east--west shifts require a transpose, shift, and transpose back, while tile transposition enables halo columns to be transferred as contiguous row segments.
They find that local data rearrangement and computation collectively cost more than halo communication.
We therefore treat within-tile transformations and arithmetic as optimization targets in the per-core realization design space.

%

\section{Mapping SIMT to a Decoupled Accelerator}
\label{sec:motivation}

\subsection{From CUDA Constructs to Blackhole}
\label{sec:pipe-reconceive}

Tenstorrent's distributed L1 memories allow a global array to be sharded across cores so that each core owns part of the array, which remains in what we call \emph{persistent L1} for the entire execution. 
We target workloads whose per-core working sets fit in local L1. 
Larger working sets additionally require scheduling DRAM transfers and managing tile reuse in L1; such planning is outside our current scope and complementary to work such as TileLoom~\cite{li2026tileloom}.
Within our scope, we follow the owner-computes rule: each output region is computed by the core that owns it. 
Affine accesses to data owned by other cores identify inter-core communication, which the compiler realizes as NoC transfers, using multicast where possible.

Once the operands have been delivered, thread-level computation must be converted into per-core tile computation. 
The current mapping associates the two CUDA thread-index dimensions with the row and column positions of a $32\times32$ tile. 
A more general mapping could first express CUDA thread and block dimensions as loops and then select two loop dimensions as the tile axes, as explored by T2T~\cite{li2026t2t}. 
The compiler then recognizes reductions and replaces them with tile-level reduction or matrix operations. 
Guards restrict the active elements through masks, while non-identical access maps require within-tile shifts, broadcasts, transposes, or boundary patches to align each operand with its consumer.

Finally, the default realization partitions the per-core program by assigning arithmetic to the compute kernel and data manipulation to the DM kernels. 
The compiler then inserts L1 buffers between producers and consumers and schedules their synchronization. 
GPUs can hide data-movement latency through warp scheduling, whereas on Blackhole the buffering and synchronization for overlap must be represented explicitly.

\subsection{Per-Core Realization Design Space}
\label{sec:motivation-two}

A DM kernel is not limited to moving data: with access to L1 and a NoC, it can also transform and compute on data through RISC-V-based scalar processing and NoC-assisted bulk copies.
The same operation can therefore be realized on DM in this way, or on compute as a tile operation.
Table~\ref{tab:reconceive} summarizes the policy dimensions exposed by the compiler.

\begin{table}[htb]
  \caption{Policy dimensions exposed by the compiler.}
  \label{tab:reconceive}
  \centering
  \footnotesize
  \setlength{\tabcolsep}{3pt}
  \renewcommand{\arraystretch}{1.15}
  \begin{tabular}{p{0.31\linewidth} p{0.61\linewidth}}
  \hline
  \textbf{Policy dimension} & \textbf{Exposed alternatives} \\
  \hline
  \noalign{\vskip 2pt}
  \multicolumn{2}{l}{\textit{C1: TilePlan}} \\
  loop scheduling & reduction blocking and loop reordering \\
  predicate placement & assignment of shared write predicates to the producing or consuming kernel \\
  \noalign{\vskip 2pt}
  \multicolumn{2}{l}{\textit{C2: TileFlow}} \\
  arithmetic placement & compute-side tile operation or DM-side scalar processing \\
  guard realization & masked select on compute, a select sunk into one operand, or a partial copy by DM \\
  axis/layout realization & direct shift/copy or transpose followed by a row-wise operation \\
  broadcast realization & full-tile replication by DM or compute-side expansion \\
  conversion placement & layout conversion before transfer at the sender or after transfer at the receiver \\
  \noalign{\vskip 2pt}
  \multicolumn{2}{l}{\textit{C3: Task}} \\
  DM-lane assignment & one DM lane or a split across DM0 and DM1 \\
  streaming granularity & synchronization per tile (column streaming), shard row (row streaming), or full shard \\
  CB depth & single or double buffering \\
  \hline
  \end{tabular}
\end{table}

A guard can be realized as a mask tile and a tile-wise select in the compute kernel, or as a DM-side copy that overwrites only the selected elements.
A single row or column used as a tile operand can be replicated across the other tile dimension by a DM kernel, or kept as a line and expanded during compute-side consumption.
Element-wise arithmetic on such a row or column need not be performed as a full-tile operation; a DM kernel can instead apply it element by element while producing the line.

The physical tile layout creates a further choice.
Row-wise shifts and copies access 16-byte-aligned contiguous segments and can use bulk NoC transfers.
Direct column-wise operations access strided elements and require repeated scalar loads and stores by a DM kernel.
Alternatively, a column-wise operation can be realized by transposing the tile and applying the corresponding row-wise operation, adding transposition overhead while enabling bulk transfers.

Streaming granularity and CB depth jointly shape the tradeoff between pipeline overlap and L1 usage.
Finer granularity may expose more overlap, but it also increases synchronization frequency and can fragment bulk transfers.

These policy choices can generally be specified independently by value and application site, while their performance effects can interact.
We expose them as declarative annotations, allowing the compiler to lower different combinations without handwritten low-level kernels.

\section{Compiler Design}
\label{sec:pipe}

\subsection{Four Compiler States and Three Policy Checkpoints}
\label{sec:pipe-overview}

\begin{figure}
  \centering
  \includegraphics[width=.8\columnwidth]{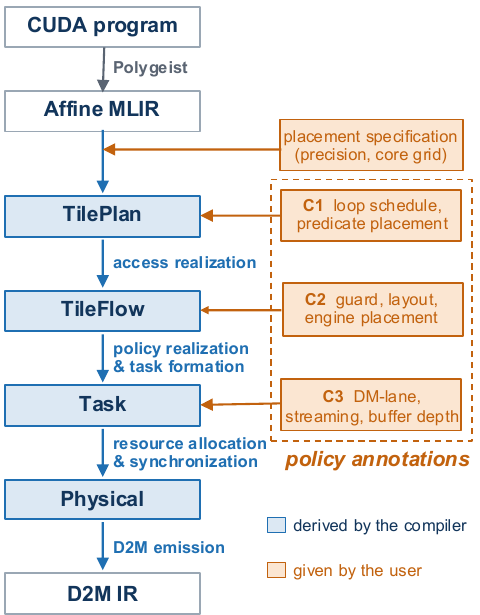}
  \caption{Four compiler states from affine MLIR to D2M IR.
  C1--C3 are policy checkpoints where users can add annotations and resume lowering. The remaining lowering is derived from the initial specification and annotations.}
  \label{fig:pipe}
\end{figure}

Figure~\ref{fig:pipe} summarizes the compiler.
It receives affine MLIR produced using Polygeist's affine-raising pass, whose iteration domains and read/write access relations enable dependence analysis.
The initial specification contains the arithmetic precision, the core grid, and tensor shapes.
It also classifies each tensor as externally visible or internal.
Lowering realizes the target program through four compiler states:

\begin{itemize}
  \item \textbf{TilePlan} --- mapping computation to cores and organizing tile-level loops, with array accesses retained as affine maps to individual elements.
  \item \textbf{TileFlow} --- explicit construction of each tile operand through cross-core transfers, local reads, within-tile shifts, masks, transposes, and broadcasts.
  \item \textbf{Task} --- operations partitioned into per-engine groups connected by logical L1 buffers, with producer--consumer synchronization placed at the selected streaming level.
  \item \textbf{Physical} --- explicit Dst register-bank assignments, absolute L1 addresses, and CB synchronization, from which the compiler emits D2M IR.
\end{itemize}

All three workloads pass through these four states without workload-specific compiler paths.
The states are not disjoint dialect boundaries: their operations may coexist, sharing the \texttt{d2mtile} dialect for tile computation.

The TilePlan, TileFlow, and Task forms serve as policy checkpoints C1--C3 in Fig.~\ref{fig:pipe}.
Table~\ref{tab:reconceive} lists the choices exposed at each checkpoint.
At each checkpoint, the user can inspect the emitted IR and attach policy attributes to operations or loops.
The compiler checks the request, applies it, and returns the transformed module to the same checkpoint.
This cycle may repeat before lowering advances to the next state.
Annotations are highlighted in blue in Listings~\ref{lst:gaussian-trace}, \ref{lst:jacobi-trace}, and~\ref{lst:streaming-trace}.
The artifact includes complete annotation sets and application steps for the reported configurations of all three workloads.

We follow the \texttt{jacobi\_2d} west operand, $W[i,j]=A[i,j-1]$, through all four states.
The \texttt{gaussian} and \texttt{syrk} examples illustrate additional policy choices.

\subsection{C1: TilePlan --- Tiled Iteration and Computation}
\label{sec:pipe-c1}

\subsubsection{Loop Scheduling}

The compiler first creates rank-2 logical views and normalizes flat loads and stores into rank-2 affine accesses.
While the program remains affine, loop-schedule annotations apply reduction blocking and loop reordering.
Applying these transformations in the Affine Dialect preserves affine accesses and guards and allows MLIR dependence analysis to validate loop reordering.
For example, in \texttt{syrk}, let $K$ denote the number of tile-level reduction steps, $B$ the block size in reduction tiles, and $S$ the number of output tiles along each dimension of one core's shard.
Listing~\ref{lst:syrk-schedule} shows the resulting C1 forms.
Strip-mining groups operand reads into $1\times B$ blocks, while interchange exposes $S\times B$ reads for both operands and keeps an $S\times S$ accumulator in L1 across reduction blocks.

\begin{lstlisting}[
  style=mlir-druvbox-light-high-small,
  float=t,
  caption={Schematic C1 schedules and logical reads for \texttt{syrk}.},
  label={lst:syrk-schedule}
]
(1) reduction innermost
shard_for (r,c)
  temporal_for (k = 0:K)
    %a,%b = logical_read %A[...] shape [1,1]
    %acc += matmul_block %a, transpose(%b) [1,1,1]

(2) strip-mine k by B
shard_for (r,c)
  temporal_for (kb = 0:K/B)
    %a,%b = logical_read %A[...] shape [1,B]
    %acc += matmul_block %a, transpose(%b) [1,1,B]

(3) interchange kb outward
temporal_for (kb = 0:K/B)
  %a,%b = logical_read %A[...] shape [S,B]
  %acc[:,:] += matmul_block %a, transpose(%b) [S,S,B]
\end{lstlisting}

\subsubsection{Automatic Lowering}

The current lowering maps the two CUDA thread-index dimensions to the row and column positions of a $32\times32$ tile.
To preserve the owner-computes rule when a CUDA kernel offsets its output indices, the compiler rebases affine accesses and guards to the destination coordinates.
For \texttt{gaussian}, this converts Rodinia's indices, which write $A[x_{\mathrm{idx}}+t+1,y_{\mathrm{idx}}+t]$, into an output-coordinate iteration space.
It then derives per-core shard loops from the tensor shapes and selected core grid.
Affine loads become logical tile reads that retain their element access maps, scalar arithmetic becomes tile operations, and stores become tile writes.
Affine conditions are represented as tile-region predicates, while sequential and reduction loops become temporal tile loops.
Host-side loops around CUDA kernel launches are lifted into device-side temporal loops, with cross-core dependences enforced later through barriers.

After TilePlan formation, the compiler fuses compatible kernels to eliminate intermediate buffers.
For example, the two CUDA kernels of \texttt{gaussian} communicate the elimination factor $m[i,t]=A[i,t]/A[t,t]$ through global memory.
Fusion forwards this value directly to the matrix update and eliminates the intermediate storage.

In Listing~\ref{lst:tileplan-c1}, the west operand of \texttt{jacobi\_2d} remains a logical read representing the global access $(i,j)\mapsto(i,j-1)$, with $i,j$ computed from core, shard-local tile, and within-tile coordinates.
Such neighbor accesses prevent fusion of the two kernels: kernel 2 would overwrite elements of $A$ that kernel 1 has not yet read.

\subsubsection{Predicate Placement Policy}

In \texttt{jacobi\_2d}, the writes in kernels 1 and 2 carry the same boundary predicates.
A C1 annotation assigns each predicate to one of the two writes and removes it from the other.
The compiler applies this policy only after verifying that the affine predicates, shard coordinates, and tile accesses are equivalent and that $B$ is an internal intermediate used only by these kernels.
Both placements preserve the boundary values of $A$.

\begin{lstlisting}[
  style=mlir-druvbox-light-high-small,
  caption={Schematic C1 form of \texttt{jacobi\_2d}.},
  label={lst:tileplan-c1}
]
// host time loop raised around both CUDA kernels
device_for (%ci, %cj) in (10, 10)
 temporal_for (%t) in (0, 1000) {
  // kernel 1: B <- stencil(A)
  shard_for (%r) axis <row> {
   shard_for (%c) axis <col> {
    %west = logical_read %A ... // west: (i,j) -> (i,j-1)
    // %b = stencil using %west and four other operands
    %g0,%g1,%g2,%g3 =
      region_predicate (i>0,i<N-1,j>0,j<N-1)
    write_if %b, %B[%r,%c]
      when(%g0,%g1,%g2,%g3) preserve %B
   }
  }
  // kernel 2: A <- B; %g0..%g3 are the same predicates
  shard_for (%r) axis <row> {
   shard_for (%c) axis <col> {
    write_if %B, %A[%r,%c]
      when(%g0,%g1,%g2,%g3) preserve %A
   }
  }
 }
\end{lstlisting}

\subsection{C2: TileFlow --- Constructing Tile Operands and Dataflow}
\label{sec:pipe-c2}

\subsubsection{Automatic Lowering}

\begin{figure}[t]
  \centering
  \includegraphics[width=\columnwidth]{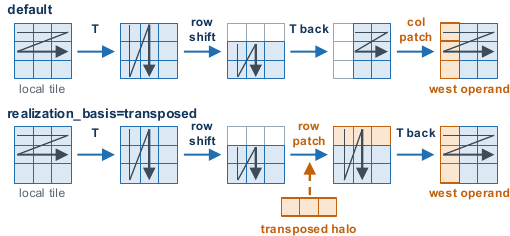}
  \caption{Two C2 realizations of the \texttt{jacobi\_2d} west operand.
  Both express the local column shift as a row shift in the transposed basis.
  The default transposes back before applying the halo as a column patch, whereas \texttt{realization\_basis=transposed} applies the transposed halo as a row patch before transposing back.}
  \label{fig:workload-realization}
\end{figure}

For each affine access, the compiler factors the array coordinates against the shard and tile geometry into a source core, a local tile coordinate, and a within-tile map.
A core map that is identically this core resolves to a local read, and one that can select another core resolves to a transfer.
The within-tile map is then normalized into the TileFlow forms shown in Listings~\ref{lst:gaussian-trace} and~\ref{lst:jacobi-trace}.
\texttt{gaussian}'s $A[t,j]$, $A[i,t]$, and $A[t,t]$ become broadcasts of a row, a column, and a cell, while \texttt{jacobi\_2d}'s west operand becomes a within-tile shift patched at the tile boundary.
Neither engine can apply a column offset efficiently, so this shift is adapted to a transpose, a row shift, and a transpose back, as illustrated in Fig.~\ref{fig:workload-realization} and Listing~\ref{lst:jacobi-trace}~(1).
Affine predicates are decomposed into row- and column-wise masks as applicable, and guarded writes become \texttt{masked\_overlay} operations that preserve the previous destination value on inactive lanes.

\subsubsection{Policies}

\paragraph{\texttt{gaussian}}
For arithmetic placement, annotating the reciprocal for DM execution makes the DM compute a scalar at the pivot cell and then broadcast it, rather than compute the reciprocal over the full tile, as shown in Listing~\ref{lst:gaussian-trace}~(2a).

For guard realization, an annotation selects a mask $m$ and a factor $x$ in a guarded update $\operatorname{select}(A-x\cdot y,A,m)$.
Under the user's assumption that $y$ is finite on lanes disabled by $m$, the compiler rewrites the update as $A-\operatorname{select}(x\cdot y,0,m)$ and sinks the select into the designated factor to obtain $A-\operatorname{select}(x,0,m)\cdot y$.
In Listing~\ref{lst:gaussian-trace}~(2b), the annotation pairs the row mask \texttt{\%mr} with the broadcast-column factor \texttt{\%col}, producing \texttt{\%masked} by zeroing inactive lanes.
A further \texttt{realization\_engine} annotation moves this fill to DM before the broadcast, producing \texttt{\%masked\_src} in step~(2c).

\begin{lstlisting}[
  style=mlir-druvbox-light-high-small,
  caption={Schematic C2 transformations of the \texttt{gaussian} update.},
  label={lst:gaussian-trace}
]
gaussian: A[i,j] -= (A[i,t] / A[t,t]) * A[t,j]
(1) Semantic TileFlow: derived access
%cell_src = tileflow.transfer %A[t,t]
%cell = tileflow.broadcast %cell_src <scalar>
%inv = d2mtile.unary recip %cell
      {tileflow.realization_engine = dm}
shard_for (%i) axis <row> {
  %col_src = tileflow.transfer %A[i,t]
  %col = tileflow.broadcast %col_src <col> from(%tc)
        {tileflow.predication_target = 0}
  %scale = d2mtile.binary mul %col, %inv
  shard_for (%j) axis <col> {
    // %row = row broadcast of A[t,j]
    %mr = tileflow.row_mask [...]  // i > t
    %mc = tileflow.col_mask [...]  // j >= t
    %out = tileflow.masked_overlay (%A[i,j]-%scale*%row)
           with %A[i,j] active(%mr,%mc)
           {tileflow.predication = <mask=0,target=0>}
  }
}

(2a) realize realization_engine on %inv
%inv_src = d2mtile.unary recip %cell_src at cell(%tr,%tc)
%inv = tileflow.broadcast %inv_src <scalar> from(%tr,%tc)

(2b) realize the predication annotations
%masked = tileflow.mask_fill %col active(%mr)
          {tileflow.realization_engine = dm}
%scale = d2mtile.binary mul %masked, %inv
%out = tileflow.masked_overlay (%A[i,j]-%scale*%row)
       with %A[i,j] active(%mc)

(2c) realize realization_engine on %masked
%masked_src = tileflow.mask_fill %col_src active(%mr)
              at col(%tc)
%masked = tileflow.broadcast %masked_src <col> from(%tc)
\end{lstlisting}

For broadcast realization, a broadcast may instead be performed as part of compute-side consumption.
The DM kernel first copies only the source line into row 0 or column 0 of a tile operand.
For example, it places \texttt{\%col\_src} in column 0, and the consuming \texttt{d2mtile.binary mul} performs the column broadcast.
For BF16, a standalone unary broadcast materializes scalar broadcasts and can also materialize row or column broadcasts when the broadcast cannot be fused into a consuming binary operation.

\paragraph{\texttt{jacobi\_2d}}
For the west operand, annotating the halo \texttt{masked\_overlay} with \texttt{realization\_basis<transposed>} requests that the halo patch be applied in the transposed basis.
This makes the patch a row copy followed by a transpose back, as illustrated in Fig.~\ref{fig:workload-realization} and Listing~\ref{lst:jacobi-trace}~(1)--(2).

An additional conversion-placement annotation selects whether the halo transpose is performed at the sender or receiver.
Under the current global-semaphore scheme, sender-side placement materializes a converted view \texttt{\%A\_t} once per time step, allowing both local reads and remote deliveries to reuse it, as shown in Listing~\ref{lst:jacobi-trace}~(3).

For guard realization, each global boundary mask is realized by default as a compute-side masked select; an annotation can move it to DM, where it becomes a conditional lane copy.
Although the C2 choice is semantically independent of C1 predicate placement, pairing compute realization with $B \leftarrow \operatorname{stencil}(A)$, or DM realization with $A \leftarrow B$, reduces compute--DM handoffs.
For each boundary mask, we therefore evaluate only these two matched placements.
\begin{lstlisting}[
  style=mlir-druvbox-light-high-small,
  caption={Schematic C2 transformation of the \texttt{jacobi\_2d} west operand.},
  label={lst:jacobi-trace}
]
jacobi west operand A[i,j-1]
(1) automatic column-shift realization + policy
%halo = tileflow.transfer %A [...]
%local = tileflow.local_read %A [...]
%local_t = tileflow.transpose %local
%shifted_t = tileflow.roll %local_t axis <row> by -1
%shifted = tileflow.transpose %shifted_t
%edge = tileflow.col_mask [...]
%west = tileflow.masked_overlay %shifted with %halo
        active(%edge)
     {tileflow.realization_basis = transposed}

(2) realize the transposed basis
%halo_t = tileflow.transpose %halo
          {tileflow.conversion_site = sender}
%west_t = tileflow.row_copy %shifted_t with %halo_t [...]
%west = tileflow.transpose %west_t

(3) realize sender-side conversion
%A_t = tileplan.converted_view %A  // holds T(A)
%halo_t = tileflow.frame_delivery %A_t [...]
%local_t = tileflow.local_read %A_t [...]
%shifted_t = tileflow.roll %local_t axis <row> by -1
%west_t = tileflow.row_copy %shifted_t with %halo_t [...]
%west = tileflow.transpose %west_t
\end{lstlisting}

\subsubsection{Derived Optimizations}
\label{sec:pipe-auto}

After applying C2 annotations, the compiler derives communication and reuse optimizations.
A transfer whose region is partly local is split into a local read and a remote delivery, and a rectangular delivery region is requested as one frame.
The compiler also identifies frame deliveries that can use multicast across a grid row or column.
For loop-level reuse, $A[t,j]$ in \texttt{gaussian} varies with the inner-loop index $j$ but not the outer-loop index $i$.
The compiler prepares a frame of the required values across $j$ before the $i$ loop and reuses it across that loop's iterations.
Transpose simplification combines $T(a)+T(b)$ into $T(a+b)$ and cancels adjacent transpose pairs.

\subsection{C3: Task --- Pipelining Compute and Data Movement}
\label{sec:pipe-c3}

After the C2 transformations, each operation is labeled as compute or DM.
A logical L1 buffer is emitted at transfer destinations and for operands constructed by DM from transferred or locally read data.
Values used only within compute remain tile values, whereas values exchanged with DM require logical L1 buffers.
In Listing~\ref{lst:streaming-trace}, \texttt{\%west\_operand} buffers \texttt{\%west\_t} from Listing~\ref{lst:jacobi-trace}~(3).
C3 also accounts for physical $16\times16$-face boundaries when materializing DM-side shifts.

The compiler then partitions operations by engine, splitting loop nests into engine-specific loops connected by the logical L1 buffers.
By default, all DM operations are assigned to DM0.
Lane annotations assign the specified operations to DM0 or DM1.
The remaining DM operations follow their dependencies when possible and otherwise fall back to DM0.
The compiler validates lane assignments to prevent cross-lane sharing of synchronized buffers and conflicting accesses to persistent L1 buffers.
In \texttt{jacobi\_2d}, the construction of the north and south operands can be assigned to one lane and that of the west and east operands to the other.
The split is legal because the north and south construction reads the original view of $A$, while the west and east construction reads the separate converted view \texttt{\%A\_t}.

\begin{lstlisting}[
  style=mlir-druvbox-light-high-small,
  caption={Schematic C3 streaming granularities for the \texttt{jacobi\_2d} west operand.},
  label={lst:streaming-trace}
]
(1) {task.streaming_granularity = <row>}
task.group @dm1 {
  %west_halo = tileflow.frame_delivery %A_t ... [0,-1]
  shard_for (%r) axis <row> {
    task.epoch_acquire %west_operand[%r]
    shard_for (%c) axis <col> {
      %shift = tileflow.roll %A_t[%r,%c] axis <row> by -1
      // %west_operand[%r,%c] = patch(%shift, %west_halo)
    }
    task.epoch_ready %west_operand[%r]
  }
}
task.group @compute {
  shard_for (%r) axis <row> {
    task.epoch_wait %west_operand[%r]
    shard_for (%c) axis <col> { stencil(%r, %c) }
    task.epoch_free %west_operand[%r]
  }
}

(2) {task.streaming_granularity = <col>}
task.group @dm1 {
  %west_halo = tileflow.frame_delivery %A_t ... [0,-1]
  shard_for (%r) axis <row> { shard_for (%c) axis <col> {
    task.epoch_acquire %west_operand[%r,%c]
    %shift = tileflow.roll %A_t[%r,%c] axis <row> by -1
    // %west_operand[%r,%c] = patch(%shift, %west_halo)
    task.epoch_ready %west_operand[%r,%c]
  }}
}
task.group @compute {
  shard_for (%r) axis <row> { shard_for (%c) axis <col> {
    task.epoch_wait %west_operand[%r,%c]
    stencil(%r, %c)
    task.epoch_free %west_operand[%r,%c]
  }}
}
\end{lstlisting}

After engine assignment, the streaming-granularity annotation selects how often the producer and consumer synchronize.
Listing~\ref{lst:streaming-trace} shows row and column streaming for the \texttt{jacobi\_2d} west operand.
Column streaming synchronizes once per tile, enabling finer-grained overlap between operand construction and computation.
Row streaming instead synchronizes once per shard row, allowing local-copy work generated by \texttt{tileflow.roll} to be batched across its columns.
If an operation sequence crosses from compute to DM and back within the column loop, as in the transpose, DM-side roll, and transpose-back sequence in Listing~\ref{lst:jacobi-trace}~(1), DM and compute must synchronize at each column-loop iteration; row streaming is therefore rejected.

\subsection{Physical --- Resource Planning and D2M Emission}
\label{sec:pipe-physical}

Physicalization resolves the logical resources introduced at C3 and emits D2M IR.
It realizes the synchronization with CB reserve/push and wait/pop operations, assigning each synchronized buffer depth one by default and depth two when streamed; a non-streamed delivery uses depth two only when explicitly annotated and it passes the compiler's legality checks.
From each buffer's shape and depth, it derives the CB capacity and allocates the L1 storage.
For \texttt{\%west\_operand} in Listing~\ref{lst:streaming-trace}, the compiler allocates L1 storage and CB capacity for one shard row per CB slot with row streaming, or one tile per slot with column streaming.
Cross-core RAW and WAR dependences on persistent L1 operands are enforced with grid-wide semaphore barriers placed before remote reads and before subsequent updates by the owner core, respectively.

For Dst allocation, the compiler places a scope inside the innermost loop that writes an output tile, assigns slots to live values by linear scan, and spills values crossing scope boundaries to L1.
A Dst-resident matmul accumulator stays within one bank across the reduction loop, whereas an L1-resident one is emitted as a \texttt{linalg.generic} whose output panel the downstream D2M pipeline divides into Dst-sized subblocks.
Each Task phase is finally emitted as \texttt{d2m.generic} over its core grid, with its DM0, DM1, and compute groups as concurrent threads.
Host--DRAM transfers, layout conversion, and DRAM--persistent-L1 movement are emitted as separate dispatches.

\subsection{Scope}
\label{sec:pipe-limitations}

The current implementation targets affine kernels over statically shaped rank-2 tensors with positive extents.
Each CUDA thread block must match the $32\times32$ tile shape.
All tensor extents must be multiples of that tile shape, and each output tile grid must divide evenly across the core grid because tail-domain schedules are not generated.
The current compiler does not support CUDA \texttt{\_\_shared\_\_} memory or \texttt{\_\_syncthreads()}.
Cross-core transfers cannot yet be split into streamed remote loads because their RAW and WAR barriers synchronize the grid around remote reads and subsequent owner updates of an entire persistent L1 buffer.

%

\section{Evaluation}
\label{sec:eval}

\subsection{Methodology}

\subsubsection{Targets}
Measurements use a Tenstorrent Blackhole p150a running system firmware \texttt{18.11.0.0} and an NVIDIA A100.
We build our backend on tt-mlir revision \texttt{2a45751}, adding LLKs for DM-side broadcasts and copies, extending the existing mask-construction LLKs to support arbitrary row and column sets, and adding lowering support for offset DMA reads and shaped remote loads.
We disable \texttt{d2m-optimize-dma} and preserve the explicit DMA and CB order emitted by our compiler.
\texttt{syrk} is compiled at math fidelity \texttt{HiFi2}; the others use the pass default, \texttt{HiFi4}.
All Blackhole configurations are compiler-generated from the scoped CUDA inputs of Section~\ref{sec:bg-workload} and run through tt-mlir, tt-metal, and ttrt.
Within each workload, they share the spatial mapping and differ only in checkpoint policies.

The Tenstorrent experiments use \texttt{gaussian} at $N=5120$ and \texttt{jacobi\_2d} and \texttt{syrk} at $N=3200$, all on a $10\times10$ core grid.
These sizes make each output tile grid evenly divisible across the core grid.
\texttt{gaussian} executes 5119 pivots, \texttt{jacobi\_2d} executes 1000 time steps, and \texttt{syrk} executes a single call.
\texttt{gaussian} and \texttt{jacobi\_2d} are evaluated at BF16 and FP32; \texttt{syrk} is evaluated at BF16 only.
\texttt{jacobi\_2d} at BF16 is numerically inaccurate after 1000 steps (Section~\ref{sec:eval-accuracy}); we report its timings only to study realization-policy effects.

For the A100 comparison, we use the same workload scope, including the omission of \texttt{gaussian}'s right-hand-side-vector update, and apply standard GPU optimizations to the Rodinia and PolyBench/ACC implementations.
For Rodinia \texttt{gaussian}, we remap the fast-varying thread index to the matrix-column dimension.
For PolyBench \texttt{jacobi\_2d}, we alternate the input and output buffers between time steps instead of copying each result back.
These GPU implementations retain the original iteration structure, without algorithmic restructuring such as blocked factorization for \texttt{gaussian} or temporal blocking for \texttt{jacobi\_2d}.
For \texttt{syrk}, we use \texttt{cublasGemmEx} with BF16 Tensor Cores and FP32 accumulation.

\subsubsection{Measurement}
We measure Tenstorrent device time with the tt-metal device profiler from the earliest kernel-zone start to the latest end across participating cores, converting cycle counts to time at 1350~MHz.
Host submission and host--device transfers are excluded.
We report the mean of five measurements per configuration.
On the A100 we additionally apply a time-based warm-up.
The coefficient of variation was at most 0.19\% across the Blackhole configurations and at most 0.6\% on the A100.

For the Tenstorrent-only policy studies, we report only the workload \texttt{d2m.generic}, with operands already resident in L1, to isolate the effects of realization policies.
For comparison with the A100, we manually fold the DRAM-to-L1 transfer and L1--DRAM writeback into the workload dispatch and report the elapsed time of the resulting single dispatch.
We exclude row-major--tiled layout conversion, assuming that Blackhole data already resides in DRAM in tiled format.
For the shorter \texttt{syrk} workload, this comparison executes 100 iterations within a single workload dispatch on Blackhole and 100 consecutive kernel launches on the A100, and we average the resulting per-iteration times.

Peak L1 is the statically planned per-core L1 footprint, including persistent L1 buffers and temporary CBs.

\subsection{Incremental Effects of Realization Policies}
\label{sec:eval-opt}

Tables~\ref{tab:gaussian-opt}--\ref{tab:syrk-opt} trace cumulative paths through the policy space.
Each row changes one policy relative to the preceding step, and the step speedup is their execution-time ratio.
Bold marks the fastest measured configuration in the ladder at each precision.
These comparisons hold the lower-level implementation fixed, including LLK code and intra-kernel scheduling.
A policy's measured effect still depends on the compiler's current lowering rules, such as Dst allocation.
Changes to either layer may change the measured effects of individual policies.
The baselines use the default realization choices except that, at C1, \texttt{jacobi\_2d} retains the boundary guards on $B\leftarrow\operatorname{stencil}(A)$.
\texttt{gaussian} uses row streaming throughout, whereas \texttt{jacobi\_2d} uses column streaming until its final BF16 row.

\subsubsection{\texttt{gaussian}}
Sinking the row mask hoists the select operation out of the column loop and improves performance at both precisions.
Sinking the column mask does not allow such a hoist and requires staging a zero tile in Dst, costing about 14\% at FP32.
At BF16, Dst allocation also changes and the subtraction runs on the FPU instead of the SFPU, holding the cost to about 1\%.
Moving the masks to DM removes mask construction and masked selection.
This change produces the largest step speedup, $3.19\times$ at BF16 and $1.93\times$ at FP32.

Moving the reciprocal to DM gains at most about 1\%, since the reciprocal is computed once per shard.
Compute-side row and column broadcasts provide further gains at BF16.
For FP32, the compute-side row broadcast is also beneficial, but adding the column broadcast increases execution time.
Instead, forming $A[i,t]/A[t,t]$ on DM provides a further $1.07\times$ at FP32 and gives the lowest measured time, as discussed in Section~\ref{sec:eval-interactions}.

\begin{table}[htb]
  \centering
  \caption{Incremental effects of realization policies for \texttt{gaussian}.}
  \label{tab:gaussian-opt}
  \scriptsize
  \setlength{\tabcolsep}{4pt}
  \begin{tabular}{lrrrr}
    \hline
    Configuration & \multicolumn{2}{c}{Time [ms]} & \multicolumn{2}{c}{Step speedup} \\
                  & BF16 & FP32 & BF16 & FP32 \\
    \hline
    base: guard as a compute select & 1290.5 & 1995.1 & -- & -- \\
    + row mask sinking & 1106.4 & 1689.6 & 1.17$\times$ & 1.18$\times$ \\
    + col mask sinking & 1119.3 & 1923.5 & 0.99$\times$ & 0.88$\times$ \\
    + masks on DM & 350.8 & 995.9 & 3.19$\times$ & 1.93$\times$ \\
    + reciprocal on DM & 347.8 & 994.9 & 1.01$\times$ & 1.00$\times$ \\
    + row broadcast on compute & 323.0 & 947.6 & 1.08$\times$ & 1.05$\times$ \\
    \hline
    \multicolumn{5}{l}{\textit{Applied separately to the row above:}} \\
    + col broadcast on compute & \textbf{309.7} & 985.4 & 1.04$\times$ & 0.96$\times$ \\
    + scale on DM & 537.8 & \textbf{889.3} & 0.60$\times$ & 1.07$\times$ \\
    \hline
  \end{tabular}
\end{table}

\subsubsection{\texttt{jacobi\_2d}}
Applying the halo patch in the transposed basis adds one tile transpose but turns a strided column patch into a contiguous row copy, providing step speedups of $1.49\times$ at BF16 and $1.29\times$ at FP32.
Moving the conversion to the sender creates a converted view that is reused by local reads and remote deliveries, reducing the number of tile transposes from five to two per output tile and providing further gains at both precisions.
Using the second DM lane provides $1.53\times$ at BF16 but little benefit at FP32, where the heavier compute stage already hides most DM work.

Moving the row boundary masks to DM has little effect, slightly slowing BF16 while giving the fastest measured FP32 configuration.
This change reduces compute-side masked selects while largely preserving contiguous $A\leftarrow B$ writeback through row-wise DM copies.
Moving the column boundary masks to DM as well removes the remaining selects from compute, and at BF16 the Dst copy they require.
However, the strided partial copies it introduces in the DM writeback cost more than the exposed saving, and both precisions slow down.

With the masks on DM, switching to row streaming provides a further $1.14\times$ at BF16, although column streaming is preferable with the masks on compute, as discussed in Section~\ref{sec:eval-interactions}.
The corresponding FP32 configuration is infeasible because it exceeds L1 capacity.

\begin{table}[htb]
  \centering
  \caption{Incremental effects of realization policies for \texttt{jacobi\_2d}.}
  \label{tab:jacobi-opt}
  \scriptsize
  \setlength{\tabcolsep}{4pt}
  \begin{tabular}{lrrrr}
    \hline
    Configuration & \multicolumn{2}{c}{Time [ms]} & \multicolumn{2}{c}{Step speedup} \\
                  & BF16 & FP32 & BF16 & FP32 \\
    \hline
    base: direct column patch & 266.8 & 414.4 & -- & -- \\
    + patch transpose & 179.7 & 321.4 & 1.49$\times$ & 1.29$\times$ \\
    + sender-side conversion & 137.6 & 201.6 & 1.31$\times$ & 1.59$\times$ \\
    + second DM lane & 89.8 & 197.0 & 1.53$\times$ & 1.02$\times$ \\
    + row masks on DM & 91.9 & \textbf{196.9} & 0.98$\times$ & 1.00$\times$ \\
    + col masks on DM & 93.1 & 200.6 & 0.99$\times$ & 0.98$\times$ \\
    + row streaming & \textbf{81.8} & -- & 1.14$\times$ & -- \\
    \hline
  \end{tabular}
\end{table}

\subsubsection{\texttt{syrk}}
Table~\ref{tab:syrk-opt} evaluates the schedule sequence in Listing~\ref{lst:syrk-schedule}.
Strip-mining groups 10 reduction tiles into shaped operand deliveries and a blocked matmul, reducing remote load transactions per core from 20,000 to 2,000 and providing $5.57\times$.
Moving the block loop outward forms larger delivery frames that are reused across the output shard, reducing the count further to 20 and providing $5.36\times$.
The larger receive buffers and an L1-resident partial accumulator increase peak L1 from 446 to 1004~KiB.
Depth-two operand delivery then provides a $1.27\times$ speedup by overlapping data transfer with computation, but raises peak L1 usage to 1404~KiB.

\begin{table}[htb]
  \centering
  \caption{Incremental effects of realization policies for \texttt{syrk} (BF16).}
  \label{tab:syrk-opt}
  \scriptsize
  \setlength{\tabcolsep}{4pt}
  \begin{tabular}{lrrr}
    \hline
    Configuration & Time [$\mu$s] & L1 [KiB] & Step speedup \\
    \hline
    base: reduction innermost & 13852.9 & 440 & -- \\
    + strip-mine $k$ & 2485.6 & 446 & 5.57$\times$ \\
    + interchange $k$ outward & 463.8 & 1004 & 5.36$\times$ \\
    + delivery depth two & \textbf{365.3} & 1404 & 1.27$\times$ \\
    \hline
  \end{tabular}
\end{table}

\subsection{Interactions among Realization Policies}
\label{sec:eval-interactions}

\begin{figure*}[t]
  \centering
  \includegraphics[width=\textwidth]{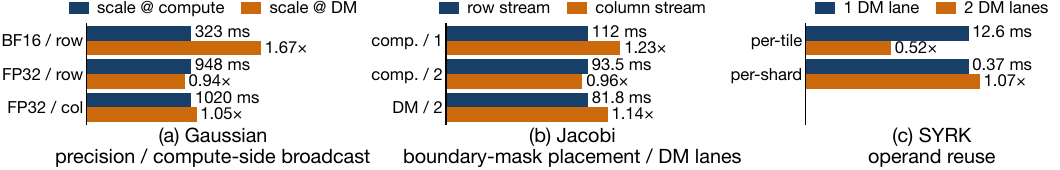}
  \caption{Realization policies whose effects reverse across surrounding configurations.
  Blue bars show absolute device time; orange bars show time normalized to the blue bar in each group.
  \texttt{gaussian} is $N=5120$; \texttt{jacobi\_2d} is $N=3200$ over 1000 steps; \texttt{syrk} is $N=3200$ at delivery depth two.}
  \label{fig:interactions}
\end{figure*}

The performance effect of a realization policy depends on the surrounding policies, arithmetic precision, and loop schedule.
Figure~\ref{fig:interactions} shows three cases in which changing this context reverses which realization is faster.

\subsubsection{Multiplication and broadcast placement in \texttt{gaussian}}
Figure~\ref{fig:interactions}(a) shows how arithmetic precision and broadcast placement affect the performance of placing the scaling multiplication $A[i,t]\cdot(1/A[t,t])$.
The scaling can be performed as a tile multiplication on compute after the column broadcast, or element-wise by DM on the source column before broadcast.

When only the row broadcast is placed on compute, DM-side scaling increases BF16 time by 67\% but reduces FP32 time by 6\%.
Under row streaming, the DM and compute stages overlap, so execution time is determined by the slower stage.
At BF16, compute is relatively light, and moving the scaling makes DM the bottleneck.
At FP32, SFPU element-wise operations make compute the bottleneck, and moving the scaling to DM relieves it.

When only the column broadcast is placed on compute, DM-side scaling instead increases FP32 time by 5\%.
A compute-side broadcast is fused into a binary operation.
With compute-side scaling, it is fused into the scaling multiplication, which is outside the column loop.
Moving the scaling to DM removes that multiplication, so the broadcast is instead fused into the other multiplication inside the column loop, increasing the number of expensive FP32 column-broadcast invocations.

\subsubsection{Streaming granularity in \texttt{jacobi\_2d}}
Figure~\ref{fig:interactions}(b) shows how the effect of column streaming depends on the DM workload.
With the boundary masks on compute, column streaming is 23\% slower with one DM lane but 4\% faster with two lanes.
With the masks on DM, it remains 14\% slower even with two lanes.
Column streaming fragments the local copies for shifted operands into smaller transfers, adding work to the DM stage.
When DM is heavy, as with one DM lane or a DM-side mask, this overhead is exposed.
With a compute-side mask and two DM lanes, compute is instead the bottleneck, so the DM work is hidden and the finer-grained pipeline improves DM--compute overlap.

\subsubsection{DM-lane assignment across \texttt{syrk} schedules}
Figure~\ref{fig:interactions}(c) shows that the benefit of assigning the multicast deliveries of $A$ and $A^T$ to separate DM lanes depends on the reduction schedule.
With the reduction innermost, splitting the fine-grained operand deliveries across two DM lanes reduces execution time by 48\% through greater transaction-level concurrency.
After interchange has reduced the delivery count to 20 per core, the same split adds coordination overhead and increases execution time by 7\%.

\subsection{Comparison with the A100}
\label{sec:eval-a100}

Table~\ref{tab:a100-comparison} compares execution time, including the device-memory traffic but excluding host--device transfers; the Blackhole side uses the fastest measured configuration at each compared precision, shown in bold in Tables~\ref{tab:gaussian-opt}--\ref{tab:syrk-opt}.

\begin{table}[htb]
  \centering
  \caption{Device-resident time excluding host--device transfers.}
  \label{tab:a100-comparison}
  \scriptsize
  \begin{tabular}{llrrr}
    \hline
    Workload & Precision & A100 [ms] & TT [ms] & TT/A100 \\
    \hline
    \texttt{gaussian}   & FP32 & 302.9 & 890.5 & 2.94$\times$ \\
    \texttt{jacobi\_2d} & FP32 & 66.3 & 197.3 & 2.98$\times$ \\
    \texttt{syrk}       & BF16 & 0.363 & 0.708 & 1.95$\times$ \\
    \hline
  \end{tabular}
\end{table}

\texttt{gaussian} and \texttt{jacobi\_2d} are $2.94\times$ and $2.98\times$ slower on Blackhole than on the A100, respectively, but the two architectures encounter different bottlenecks.
Even at the A100's theoretical peak DRAM bandwidth of 1.555~TB/s, the DRAM traffic measured with Nsight Compute would require at least 68\% of \texttt{gaussian} execution time and 79\% of \texttt{jacobi\_2d} execution time, showing that DRAM traffic alone accounts for most of their execution time.

\begin{figure}[tb]
  \centering
  \includegraphics[width=\columnwidth]{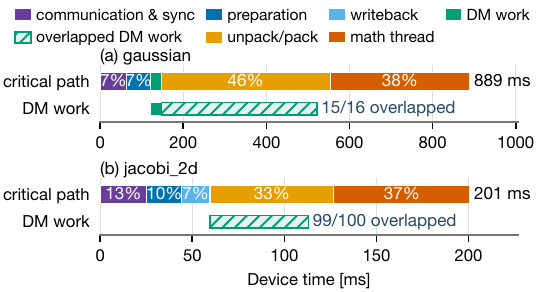}
  \caption{Estimated breakdown of FP32 Blackhole execution time.
  Preparation denotes DM-side pivot-row/cell construction in \texttt{gaussian} and compute-side transposed view construction in \texttt{jacobi\_2d}.
  The hatched portions of DM work denote pivot-column construction and scaling in \texttt{gaussian} and halo construction in \texttt{jacobi\_2d}, estimated by assuming that only the first of 16 shard rows or 100 tiles, respectively, is exposed.}
  \label{fig:blackhole-breakdown}
\end{figure}

Unlike the A100 implementations, the Blackhole realizations keep their working sets in distributed L1.
Figure~\ref{fig:blackhole-breakdown} estimates an execution-time breakdown from controlled variants.
Differences measure the incremental cost of restoring each mechanism under the corresponding configuration.
The \texttt{jacobi\_2d} decomposition uses the all-boundary-masks-on-DM configuration rather than the fastest measured configuration to separate stencil arithmetic cleanly.
Cross-core communication and CB synchronization account for only 7\% of \texttt{gaussian} and 13\% of \texttt{jacobi\_2d} execution time and do not dominate either workload.
Serialized preparation adds another 7\% and 10\%, and \texttt{jacobi\_2d}'s $A\leftarrow B$ writeback a further 7\%.
Most DM-side tile construction is hidden behind compute, showing that the generated pipeline effectively overlaps DM and compute work.

The remaining bottleneck lies in operand staging and SFPU execution within the compute kernel.
In the evaluated FP32 path, unpack-to-Dst transfers and SFPU operations serialize on the math thread.
Each SFPU operation sequentially executes arithmetic instructions, loads and stores between Dst and the SFPU's local registers, and address updates.
These costs limit throughput below the arithmetic peak and motivate further investigation of intra-kernel scheduling and optimization.

For \texttt{syrk}, the L1-resident Blackhole workload dispatch sustains 179~TFLOP/s, or 65\% of the 276~TFLOP/s peak scaled to the 100 evaluated cores.
Including the serialized DRAM--L1 transfers increases execution time from $0.365$ to $0.708$~ms.
The A100 completes the same device-resident workload, including device-memory accesses, in $0.363$~ms, also sustaining 180~TFLOP/s, or 58\% of its 312~TFLOP/s BF16 peak.
The L1-resident Blackhole execution therefore achieves a fraction of peak performance comparable to that of the A100's end-to-end execution, while overlapping DRAM--L1 transfers with the contraction remains the primary opportunity on Blackhole.

\subsection{Numerical Accuracy}
\label{sec:eval-accuracy}

We report relative Frobenius error for the fastest configuration at each precision, against an FP64 reference computed in PyTorch from the same inputs.
For \texttt{gaussian} on a diagonally dominant input, the FP32 error is $9.3\times10^{-7}$ over the off-diagonal; for \texttt{jacobi\_2d}, it is $9.5\times10^{-6}$ over the interior after 1000 steps.

At BF16, the corresponding errors increase to $4.7\times10^{-3}$ for \texttt{gaussian} and $1.68$ for \texttt{jacobi\_2d}.
Both device outputs are nevertheless reproduced bit-exactly, at their full evaluated depths, by a simulator that models the emitted operation order, the FPU/SFPU assignment, the BF16 constants, and the accumulation of multiplier partial products in BF16 Dst with \texttt{fp32\_dest\_acc\_en=false}.
The increase therefore reflects BF16 representation and Tensix arithmetic; in particular, $0.2$ is not exactly representable in BF16, introducing a systematic bias at every \texttt{jacobi\_2d} step.
For \texttt{syrk} at BF16, the whole-output error is $1.0\times10^{-2}$; a simulator of the measured matrix-unit arithmetic likewise reproduces the device output bit-exactly.
Future annotations could selectively promote operation and Dst-accumulation precision, trading accuracy against Dst capacity and execution time.

\subsection{Toward Feedback-Driven Annotation Selection}
\label{sec:eval-feedback}

The annotation space is large because policy choices can be combined and, where applicable, specified independently for different axes.
Even without varying CB depth, the evaluated BF16 space contains 336 configurations for \texttt{gaussian} and 256 for \texttt{jacobi\_2d}, all of which executed successfully.
These conservative counts exclude unlikely but legal choices, such as sinking masks along unrelated axes or adding more compute--DM handoffs.
The performance differences in Tables~\ref{tab:gaussian-opt}--\ref{tab:syrk-opt} and the policy interactions in Fig.~\ref{fig:interactions} motivate exposing and comparing alternative per-core realizations rather than relying on fixed per-annotation costs.

We therefore envision a feedback-driven system in which an LLM proposes annotation sets and uses measured performance to guide subsequent proposals, while the compiler applies each set and completes the lowering.
Structural legality checking and low-level code generation remain within the compiler, while the LLM searches the declarative annotation space.

\section{Related Work}
\label{sec:related}

Prior work retargets GPU programs across architectures.
MKPipe~\cite{liu2020mkpipe} pipelines OpenCL multi-kernel workloads on FPGAs and balances throughput and resources across kernels, while Moses et al.~\cite{moses2023high} translate CUDA parallel constructs to CPUs.
Booth~\cite{booth} accepts CUDA, HIP, and Triton and emits Tenstorrent Metalium through a Tile DataFlow IR that represents regions, communication channels, transfer and synchronization actions, and L1 placement.
HetGPU~\cite{yang2025hetgpu} instead proposes a compiler and runtime layer for cross-vendor binary compatibility by mapping CUDA SIMT semantics to heterogeneous GPUs, including Tenstorrent's many-core MIMD execution model.
Our work instead focuses on affine CUDA HPC kernels and preserves flexibility in their per-core realization through declarative annotations.
T2T~\cite{li2026t2t} and TileLoom~\cite{li2026tileloom} address complementary thread-to-tile mapping and spatiotemporal dataflow planning, respectively, as discussed in Section~\ref{sec:motivation}.

Accelerator-oriented programming systems instead expose spatial execution and data movement directly to the programmer.
TT-Lang~\cite{ttlang} provides a Python DSL for writing Tenstorrent kernels that progressively discloses pipelining and synchronization details.
Dato~\cite{fang2025dato} similarly provides a task-based model for FPGA and NPU dataflow accelerators, in which tasks are connected through explicit streams and input sharding is expressed as layout types.
In contrast, our compiler starts from existing CUDA code rather than a new DSL, and derives inter-core communication from its affine accesses rather than requiring explicit tasks and streams as in Dato.

ARIES~\cite{zhuang2025aries} and Tiramisu~\cite{baghdadi2019tiramisu} provide user-directed schedules for choices such as computation placement and loop transformations.
Our annotations instead target fine-grained per-core realization choices for decoupled compute and data-movement processors, including operation placement, guard realization, and streaming granularity.
TT-Lang is also designed to support AI-assisted translation and optimization in which AI agents generate and refine programs in its DSL, whereas Merouani et al.~\cite{merouani2025agentic} use an LLM to propose Tiramisu transformation sequences whose legality and performance are evaluated by the compiler.
Our checkpoints could similarly support LLM-proposed annotations, with the compiler checking legality and performing low-level lowering.

\section{Conclusion}
We presented an MLIR-based compiler that derives spatial mapping and per-core realization when migrating affine CUDA programs to Tenstorrent Blackhole, exposing per-core realizations as declarative annotations and lowering their combinations automatically.
The fastest measured per-core configurations improve performance by 4.2$\times$ on BF16 \texttt{gaussian} and 2.1--2.2$\times$ on FP32 \texttt{gaussian} and \texttt{jacobi\_2d}, relative to the compiler's own default realizations.
The measurements also show that the effect of an annotation depends on the surrounding policies.

\section*{Acknowledgment}
This work is supported in part by JST CREST JPMJCR21D2.
The authors used LLM assistants (Claude and Codex) for drafting and refining text from author-provided outlines and technical notes, and for coding support.

\bibliographystyle{IEEEtran}
\bibliography{ref}

\clearpage
\section*{Appendix: Artifact Description/Artifact Evaluation}
\label{sec:ad}

\subsection*{Artifact Description (AD): Artifact Identification}
The artifact supports the realization policies of Section~\ref{sec:pipe} and the measurements of Section~\ref{sec:eval}.
It is available at \url{https://github.com/casys-utokyo/annotation-driven-cuda-to-tenstorrent} and includes four main components.

\begin{itemize}
  \item \textbf{Tenstorrent backend fork}: \url{https://github.com/casys-utokyo/tt-mlir}, branch \texttt{waccpd26}, based on upstream \texttt{2a45751}, containing the LLK and lowering extensions described in Section~\ref{sec:eval}.
  \item \textbf{MLIR snapshots}: CUDA-derived inputs and intermediate representations through final D2M, organized by configuration in directories corresponding to the incremental-policy tables, interaction figure, and A100 comparison.
  \item \textbf{Compiled Blackhole programs}: each executable configuration's measured \texttt{80-program.ttm} and the shared \texttt{system\_desc.ttsys}. The flatbuffers contain the generated kernels and run with upstream ttrt.
  \item \textbf{A100 sources}: the comparison implementations in \texttt{a100-hand-optimized/}, using ping-pong buffers for Jacobi, coalesced accesses for Gaussian, and cuBLAS for SYRK.
\end{itemize}

The artifact supports inspection of the compiler representations, execution of the measured programs, and recompilation starting from \texttt{70-final-d2m.mlir} using the backend fork.
It does not include the CUDA-to-D2M frontend, so it cannot regenerate these snapshots from CUDA or apply new frontend annotations.
Each configuration's \texttt{20-policy-summary.md} describes its annotations and target operations, while \texttt{REPRO.sh} records the annotation application and lowering sequence.

\subsection*{Artifact Evaluation (AE): Reproducibility of Experiments}
\paragraph{Requirements}
Executing Blackhole programs requires a p150a and upstream ttrt at tt-mlir revision \texttt{2a45751}.
Recompiling final D2M requires the backend fork.
Accuracy checks additionally require PyTorch and NumPy.
The A100 experiments use an NVIDIA A100-PCIE-40GB with driver \texttt{560.35.05} and CUDA Toolkit~12.0.

\paragraph{Workflow}
\begin{enumerate}
  \item Select a configuration corresponding to Tables~\ref{tab:gaussian-opt}--\ref{tab:a100-comparison} or Fig.~\ref{fig:interactions} using the repository README. Each configuration directory contains its policy documentation and compiler snapshots.
  \item Reproduce device timings with \texttt{tools/timing/zones.py}, following \texttt{tools/timing/README.md} for profiler setup and workload-specific options. One measurement with this tool takes about half a minute.
  \item Reproduce numerical errors and BF16 simulator comparisons with \texttt{check.py} and \texttt{emulate.py} in \texttt{tools/accuracy/}, following its README for the configurations in Section~\ref{sec:eval-accuracy}.
  \item Build and run \texttt{a100-hand-optimized/} following its README to reproduce the A100 comparison.
\end{enumerate}

\paragraph{Expected results}
Measured times should agree with the corresponding table rows and figure bars within experimental variation.
The accuracy tools report relative Frobenius errors comparable to Section~\ref{sec:eval-accuracy} and bit-exact agreement with the BF16 simulator.

\end{document}